\documentclass[aip,graphicx, preprint]{revtex4-1}

\draft % marks overfull lines with a black rule on the right

\usepackage{graphicx}% Include figure files
\usepackage{dcolumn}% Align table columns on decimal point
\usepackage{bm}% bold math
\usepackage[utf8]{inputenc}
\usepackage[T1]{fontenc}
\usepackage{mathptmx}
\usepackage{xcolor}
\usepackage{amsmath}

\begin{document}

\title{Reduction of surface spin-induced electron spin relaxations in nanodiamonds} %Title of paper

\author{Zaili Peng}
\affiliation{Department of Chemistry, University of Southern California, Los Angeles, California 90089, USA}

\author{Jax Dallas}
\affiliation{Department of Chemistry, University of Southern California, Los Angeles, California 90089, USA}

\author{Susumu Takahashi}
\email{susumu.takahashi@usc.edu}
\affiliation{Department of Chemistry, University of Southern California, Los Angeles, California 90089, USA}
\affiliation{Department of Physics \& Astronomy, University of Southern California, Los Angeles, California 90089, USA}

\date{\today}

\begin{abstract}
Nanodiamonds (NDs) hosting nitrogen-vacancy (NV) centers are promising for applications of quantum sensing. 
Long spin relaxation times ($T_1$ and $T_2$) are critical for high sensitivity in quantum applications. 
It has been shown that fluctuations of magnetic fields due to surface spins strongly influences $T_1$ and $T_2$ in NDs. However, their relaxation mechanisms have yet to be fully understood. 
In this paper, we investigate the relation between surface spins and $T_1$ and $T_2$ of single-substitutional nitrogen impurity (P1) centers in NDs.
The P1 centers located typically in the vicinity of NV centers are a great model system to study the spin relaxation processes of the NV centers. 
By employing high-frequency electron paramagnetic resonance (EPR) spectroscopy, we verify that air annealing removes surface spins efficiently and significantly reduces their contribution to $T_1$.
\end{abstract}

\maketitle %\maketitle must follow title, authors, abstract and \pacs

\section{Introduction}
Diamond is a fascinating material in physics, chemistry and biology. 
For example, a negatively charged nitrogen-vacancy (NV) center in diamond is a promising platform for fundamental sciences and applications of quantum sensing because of its unique magnetic and optical properties as well as a long coherence time at room temperature.~\cite{Gruber1997, Wrachtrup2006, Epstein05, Gaebel06, Childress06, Takahashi08, Degen08, Maze2008, Balasubramanian08, Hall2009, Maletinsky2012}
Magnetic sensing using a single NV center has been utilized to improve the sensitivity of electron paramagnetic resonance (EPR) spectroscopy to the level of a single spin.~\cite{deLange12, Grinolds13, Tetienne2013, Mamin2013, Grinolds14, Kaufmann2013, Sushkov2014, Shi2015, Abeywardana16}
NV-detected EPR allows the detection of external spins existing around the NV center within several nanometers. 
NV-based sensing is also useful to detect electric field, temperature, strain and pH value in a nanoscale volume.~\cite{Dolde11, Jarmola12, Cai14, Fujisaku2019}
In NV-detected magnetic sensing, a magnetic field is detected through the measurement of the spin relaxation times of NVs such as $T_2$ and $T_1$. 
For example, in NV-based AC magnetic sensing measurement using a spin echo sequence, the detectable magnetic field is proportional to $1/\sqrt{T_2}$.~\cite{Taylor08}
A small number of Gd$^{3+}$ spins has been detected through sensing of fluctuating magnetic fields from Gd$^{3+}$ spins.~\cite{Tetienne2013}
In this case, the detection is achieved by measuring changes of $T_1$ relaxation time and the detectable magnetic field is proportional to $1/T_1$. 
Thus, long $T_1$ and $T_2$ times are desired for high detection sensitivity.

In NV-based magnetic sensing applications, it is also critical to position the NV center near a target of the magnetic field sensing. 
NVs located near the diamond surface and NVs in nanodiamonds (NDs) will therefore be an ideal platform for the applications.
However, $T_1$ and $T_2$ relaxation times of those NVs are often significantly reduced by surface defects and impurities including dangling bonds, graphite layers and transition metals.~\cite{De2007, Tisler2009, Kaufmann2013, Song2014, Myers2014, Rosskopf2014, Ofori2012, Tetienne2013, Iakoubovskii2000, Shames2002, Soltamova2009, Dubois2009, Peng2019}
For instance, it has been reported that shorter $T_1$ and $T_2$ were observed from shallow NVs.~\cite{Ofori2012, Myers14}
It has also been reported that $T_1$ of NVs in NDs is shorter than $T_1$ in bulk diamond. 
The recent study showed that $T_1$ of NV centers is shorter in a smaller size of NDs and the result implies a decoherence process due to surface impurities although the surface impurities were not measured in the study.~\cite{Tetienne2013}
Moreover, control of the diamond surface enables the determination of spin relaxation mechanisms, subsequently improving the sensitivity of the NV-based magnetic sensing techniques. The recent experiment by Tsukahara et al. showed that air annealing efficiently removes graphite layers compared with tri-acid cleaning and increases the $T_2$ time 1.4 times longer.~\cite{Tsukahara2019}

In this paper, we investigate the relation between surface spins and $T_1$ and $T_2$ of single substitutional nitrogen impurity (P1) centers in NDs using high-frequency (HF) EPR spectroscopy.
Our previous study on NDs suggested that the surface spins are dangling bonds located in the surface shell with a thickness of $\sim 9$ nm.~\cite{Peng2019}
Therefore, the present study aims to remove the surface spins by etching of NDs more than 9 nm and improve the spin relaxation times. 
Although $T_1$ and $T_2$ of NV centers are the primary interest for the quantum sensing applications, there are a few advantages to study the spin relaxation on P1 centers over NV centers. 
First, NV centers are located near P1 centers, shown by the detection of their magnetic dipole coupling via double electron-electron resonance spectroscopy.~\cite{deLange12, Abeywardana16, Stepanov16}
Therefore, their $T_1$ and $T_2$ times are similar and the relaxation mechanisms are often common.~\cite{Takahashi08}
Second, as shown in the previous study,~\cite{Peng2019}
EPR signals of both P1 and surface spins are observable in the same measurement. 
This allows us to determine the amount of surface spins and to study the spin relaxations using the same ND samples.
In the experiment, we employ air annealing to etch the diamond surface efficiently. 
The performance of the air annealing is confirmed by dynamic light scattering (DLS) and 230 GHz EPR experiments. 
The result of the DLS characterization shows a uniform etching and a linear etching rate of ND samples. We also confirm the reduction of the surface spins after the annealing process with high resolution 230 GHz EPR spectral analysis. 
Then, we investigate $T_1$ of P1 centers after the annealing using 115 GHz pulsed EPR spectroscopy. 
The 115 GHz EPR configuration is advantageous for pulsed EPR experiment because of its higher output power. 
The temperature and size dependence study elucidates surface spin-induced $T_1$ process. 
From the result, we find that air annealing significantly reduces the presence of surface spins, but a small fraction remains,
even after the thickness of NDs is reduced more than 9 nm. 
We also find that the surface spin contribution on $T_1$ is suppressed by a factor of $7.5 \pm 5.4$ after annealing at 550 $^{\circ}$C for 7 hours.
With the same annealing condition, $T_2$ is improved by a factor of $1.2 \pm 0.2$.

EPR signals of both P1 and surface spins are observable in the same measurement.
This allows us to determine the amount of surface spins and to study the spin relaxations using sample samples.
In the experiment, we employ air annealing to etch the diamond surface efficiently.
The performance of the air annealing is confirmed by dynamic light scattering (DLS) and 230 GHz EPR experiments.
The result of the DLS characterization shows a uniform etching and a linear etching rate of ND samples.
We also confirm the reduction of the surface spins after the annealing process with high resolution 230 GHz EPR spectral analysis.
Then, we investigate $T_1$ of P1 centers after the annealing using 115 GHz pulsed EPR spectroscopy.
The 115 GHz EPR configuration is advantageous for pulsed EPR experiment because of its higher output power.
The temperature and size dependence study elucidates surface spin-induced $T_1$ process.
From the result, we find that air annealing significantly reduces the presence of surface spins, but a small fraction remains, even after the thickness of NDs is reduced more than $9$ nm.
We also find that the surface spin contribution on $T_{1}$ is suppressed by a factor of $7.5 \pm 5.4$ after annealing at 550 $^{\circ}$C for 7 hours.
With the same annealing condition, $T_2$ is improved by a factor of $1.2 \pm 0.2$.

\section{Materials and Methods}
\subsection{Nanodiamond}
Five different sizes of diamond powders were investigated in the present study.
The samples include micron-sized diamond powders ($10\pm 1$ $\mu$m) (Engis Corporation), and four different sizes of NDs (Engis Corporation and L.M. Van Moppes and Sons SA).
The mean diameters of the ND samples specified by the manufacturers are $550\pm100$ nm, $250\pm80$ nm, $100\pm30$ nm, and $50\pm20$ nm.
All diamond powders were manufactured by mechanical milling or grinding of type-Ib diamond crystals.
The concentration of nitrogen related impurities in the ND powders is in the order of 10 to 100 parts per million (ppm) carbon atoms.

\subsection{Air annealing}
The air annealing process was performed using a tube furnace (MTI Corporation) where a ND sample was positioned in a quartz tube located in the cylindrical access of the furnace.
For the preparation of the annealing process, the ND sample was placed in a 5 ml of acetone.
The ND sample in acetone was then mixed by utilizing ultrasound sonication for 10 min at room temperature in order to achieve uniform dispersion.
After the ultrasound sonication, the sample solution was placed in a crucible and kept in a fume hood overnight (without application of heating) in order to evaporate acetone from the crucible.
In the air annealing process, the temperature of the furnace was first stabilized at the annealing temperature (550 $^\circ$C in the present case), and then the ND sample in the crucible was inserted at the center of the quartz tube.
In order to improve homogeneity of the application of the air annealing over the ND powders, the NDs were mixed by a lab spatula periodically during the annealing (typically mixed for 30 seconds every 10 minutes).
We also limited the initial amount of ND samples to be approximately 30 mg for the homogeneous application of the air annealing.

\subsection{Dynamic light scattering}
The size of a diamond powder sample was characterized by dynamic light scattering (DLS) (Wyatt Technology).
A diamond powder sample of $\sim 1$ mg was suspended in methanol and sonicated for two hours before the measurement of DLS.
The DLS measurement was performed with a 632 nm incident laser and $163.5^{\circ}$ of detection angle.
The second correlation data was analyzed using the constrained regularization method to obtain particle sizes.

\subsection{HF EPR spectroscopy}
HF (230 GHz and 115 GHz) EPR experiments were performed using a home-built system at University of Southern California.
The HF EPR spectrometer consists of a high-frequency high-power solid-state source, quasioptics, a corrugated waveguide, a 12.1 Tesla superconducting magnet, and a superheterodyne detection system.
The output power of the source system is 100 mW at 230 GHz and 480 mW at 115 GHz, respectively.
A sample on a metallic end-plate at the end of the corrugated waveguide is placed at the center of the 12.1 Tesla EPR superconducting magnet.
Details of the system have been described elsewhere.~\cite{Cho2014}
In the present study, the diamond powder sample was placed in a Teflon sample holder (5 mm diameter), typically containing the diamond powders of 5 mg.~\cite{Cho2015}
For cw EPR experiments, the microwave power and magnetic field modulation strength were adjusted to maximize the intensity of EPR signals without distortion of the signals.~\cite{Peng2019}
A typical modulation amplitude was 0.02 mT at a modulation frequency of 20 kHz.

\section{Results and Discussion}
\begin{figure}
\includegraphics [width=60 mm]{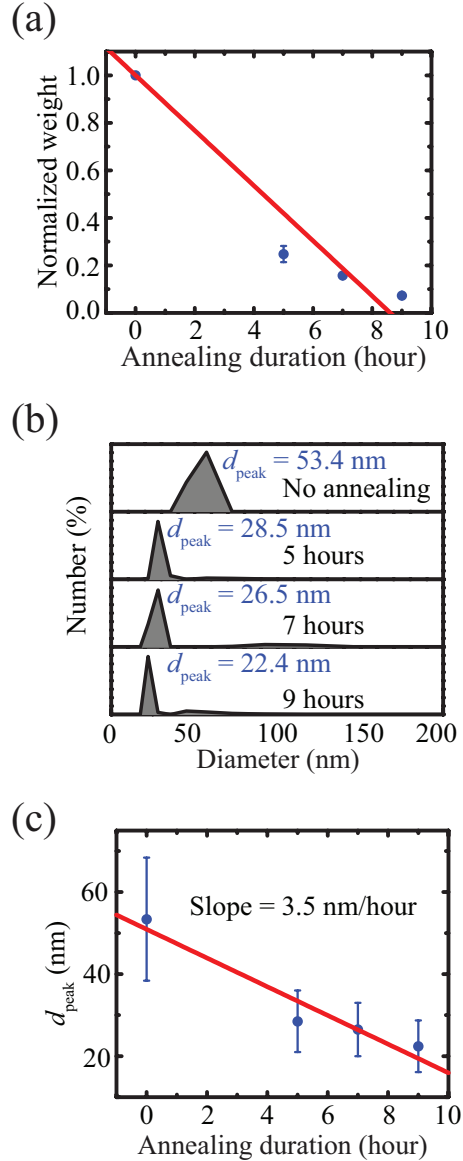}
\caption{\label{fig:AirA} Overview of the air annealing experiment.
(a) The normalized weight as a function of the annealing duration with annealing at 550 $^{\circ}$C.
The red solid line shows a linear fit to obtain the rate of weight reduction.
The weight reduction rate was $0.12$ hour$^{-1}$.
Each sample was weighed five times.
The error bar represents the standard deviation of the measurements.
(b) DLS results for the size characterization of the ND samples before and after the annealing for 5, 7 and 9 hours.
The diameter at the maximum in the distribution ($d_{peak}$) is indicated.
The obtained polydispersity index (PDI) were 0.11, 0.07, 0.06 and 0.08 for the no-annealing sample and the annealing for 5, 7 and 9 hours samples, respectively.
(c) $d_{peak}$ as a function of the annealing duration.
The red solid line represents the result of a linear fit.
The error bar represent the standard deviation (calculated by $d_{peak}\sqrt{PDI}$).
}
\end{figure}
We employed air annealing for the removal of the surface spins in the present study.
In the air annealing the surface removal is caused by etching by oxygen where oxygen molecules oxidize carbon and form gaseous CO and CO$_2$.
We first compared the weight of the ND sample before and after the annealing process.
Figure~\ref{fig:AirA}(a) shows the ND normalized weight as a function of the annealing time.
In the experiment, the annealing was done at an annealing temperature of 550 $^{\circ}$C.
The result shows linear reduction in ND weight with increased annealing time.
The size of the ND samples was then characterized using DLS.
As shown in Fig.~\ref{fig:AirA}(b), the ND size decreased from $d_{peak} = 53.4$ nm to 22.4 nm after the annealing for 9 hours.
The observed reduction and narrow distribution of the size indicates a successful and uniform application of the annealing to NDs.
Figure~\ref{fig:AirA}(c) shows the ND size as a function of the annealing duration.
We observed a linear relationship between the size reduction and the annealing duration.
A reduction rate of $3.5$ nm/hour was obtained from the linear fit.
\begin{figure}
\includegraphics [width=70 mm]{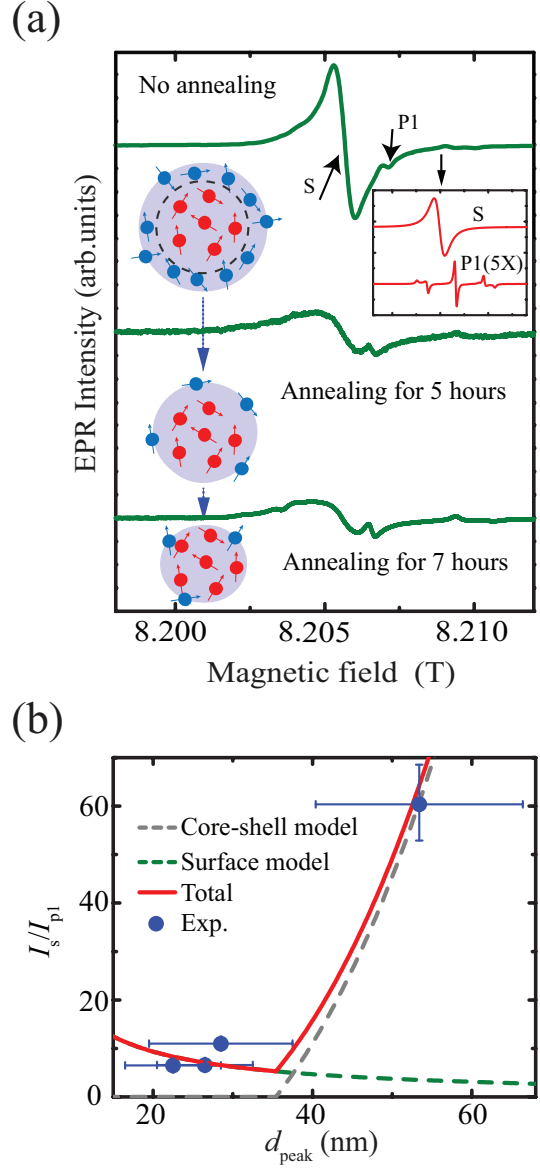}
\caption{\label{fig:EPRspec}
cw EPR analysis of 50-nm NDs before and after the air annealing.
(a) Signal intensity as a function of magnetic fields in Tesla with no annealing, annealing for 5 hours and 7 hours.
The solid green lines represent the experimental data.
The inset on the top right shows contributions of P1 and surface spins (S) on the EPR spectrum, which were extracted from the EPR spectral analysis.
Drawings representing NDs under the annealing process are also shown in the inset.
The red arrows in the drawing represent the P1 centers, and the blue arrows represent surface spins.
(b) The EPR intensity ratio $I_{s}/I_{P1}$ as a function of the diameter ($d_{peak}$).
The blue solid circles with error bars represent $I_{s}/I_{P1}$ obtained from EPR spectral analysis.
The details of the EPR spectral analysis is described in Supplementary Material.
The gray dashed line is the simulated $(I_{s}/I_{P1})_{coreshell}$.
The green dashed line is the simulated $(I_{s}/I_{P1})_{surface}$.
}
\end{figure}

Next, we characterized paramagnetic spins existing in NDs using 230 GHz EPR spectroscopy.
Figure~\ref{fig:EPRspec}(a) shows 230 GHz continuous-wave EPR spectra on 50-nm ND samples before and after the air annealing.
The measurements were performed at room temperature.
As shown in Fig.~\ref{fig:EPRspec}(a), all spectra contain a pronounced and broad EPR signal at 8.206 Tesla and a narrow EPR signal at $\sim$8.207 Tesla. From the EPR spectral analysis shown in the inset of Fig.~\ref{fig:EPRspec}(a), we identified that the EPR signal at 8.207 Tesla is from P1 centers while the signal at 8.206 Tesla is from the surface spins (dangling bonds).
The result is consistent with the previous HF EPR study.~\cite{Peng2019}
As shown in Fig.~\ref{fig:EPRspec}(a), the intensity of the EPR signals from the surface spins decreases significantly after the annealing.
In general, the EPR intensity is related to the spin population, we therefore analyzed the EPR intensity ratio between the surface spins ($I_S$, where S represents surface spins) and P1 ($I_{P1}$) to determine their spin population ratio.
For example, we obtained $I_{S}/I_{P1}$ to be 61 and 5 with no annealing and 9 hour annealing, respectively.
The result from the EPR intensity and DLS analyses was summarized in Fig.~\ref{fig:EPRspec}(b).
Since our previous HF EPR study of the non-annealed NDs showed the core-shell structure with the shell thickness ($t$) of 9 nm,~\cite{Peng2019} we first consider the core-shell model to understand the size dependence of the EPR intensity.
In the core-shell model, the EPR intensity ratio,
\begin{equation*}
\left( \frac{I_X}{I_{P1}} \right)_{coreshell} = \frac{\rho_X V_X}{\rho_{P1} V_{P1}} = \frac{\rho_X [4\pi/3 \{(d/2)^3-(d/2-t)^3 \} ] }{\rho_{P1} [4\pi/3(d/2-t)^3]},
\end{equation*}
where $\rho_{X}$ ($\rho_{P1}$) is the density of the surface spins (P1 spins) and $V_X$ ($V_{P1}$) is the volume of the surface spin (P1 spin) locations.
The calculated $(I_S/I_{P1})_{coreshell}$ is shown in Fig.~\ref{fig:EPRspec}(b).
However, we observed a poor agreement with the experimental data in the range of $d < 35$.
There may be two possible reasons to explain the result.
First, as reported previously~\cite{Gaebel2012, Wolfer2009, De2000, Dallek1991, Xie2018}, the etching rate of the air annealing depends on a crystallographic axis. 
It has been shown that the etching rate of the (111) plane is a couple of times faster than the (100) plane.~\cite{Sun1992}
However, this can explain only the dependence of EPR, but not the dependence of DLS.
Another possible reason is the creation of a small amount of surface spins during air annealing.
For instance, it has been reported that dangling bonds were created by air annealing, especially when the surface termination was dominated by C-H bonds.~\cite{Wenjun1992}
In the latter scenario, the surface spins in the non-annealed NDs (dangling bonds) are located in the shell, and then air annealing removes the dangling bonds in the shell as well as creates a small amount of dangling bonds on the surface (see Fig.~\ref{fig:EPRspec}(a)).
To take into account the surface spins created by air annealing, we added a contribution from the surface spin model with which,
\begin{equation*}
\left( \frac{I_X}{I_{P1}} \right)_{surface} = \frac{\rho_s [4\pi (d/2)^2]}{\rho_{P1} [4\pi/3 (d/2)^3)]} \propto \frac{\rho_s}{d}.    
\end{equation*}
$\rho_s$ is the surface spin density, treating as a constant here.
As shown in Fig.~\ref{fig:EPRspec}(b), the sum of the core-shell and surface models agrees with the experimental result, supporting the latter case.

\begin{figure}
\includegraphics [width=50 mm]{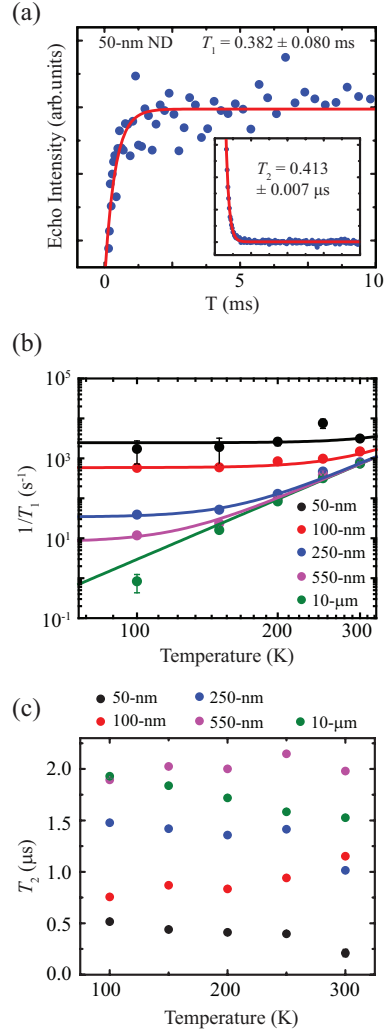}
\caption{\label{fig:T1} Temperature dependence of $T_{1}$ and $T_2$ of P1 centers in NDs.
(a) The $T_1$ measurement using the inversion recovery measurement.
The measurement was performed at 200 K.
The pulse sequence is $P_{\pi}-T-P_{\pi/2}-\tau-P_{\pi}-\tau-$echo where $P_{\pi/2}$ and $P_{\pi}$ are $\pi/2$- and $\pi$-pulses, respectively, $\tau$ is a fixed evolution time and an evolution time $T$ is varied in the measurement.
In the measurement, the pulse lengths of $P_{\pi/2}$ and $P_{\pi}$ were 300 ns and 500 ns, $\tau = 1.2$ $\mu$s and the repetition time was $=10$ ms.
The inset shows the spin echo measurement at 200 K to obtain $T_2$.
The pulse sequence is $P_{\pi/2}-\tau-P_{\pi}-\tau-$echo where $\tau$ is varied in the measurement.
The pulse parameters for the $T_2$ measurement were $P_{\pi/2} = 300$ ns, $P_{\pi} = 500$ ns and the repetition time = 10 ms.
The errors associated with $T_1$ and $T_2$ were obtained by computing the standard error.
(b) Temperature dependence of $1/{T_{1}}$ on various sizes of NDs.
The solid circles are experimental data and the solid lines are fits using Eq.~(\ref{rate}).
(c) $T_2$ on various sizes of NDs.
The error bars are smaller than the dots representing the $T_2$ value.
}
\end{figure}
\begin{table}[htb!]
\caption{\label{tab:T1} Summary of $T_1$ analyses. For the $\Gamma_s$ analysis, Eq.~\ref{rate} and $C = 2.96 \times 10^{-10}$ ($s^{-1}K^{-5}$) were used.
$T_1$ and $\Gamma_s$ are shown with three significant figures.
The errors in $T_1$ represent the standard error of the mean.
The errors in $\Gamma_s$ were calculated as the 95\% confidence interval.}
\begin{center}
\begin{tabular} {|c |c |c |c |c |c |c | }
\hline
Sample & \multicolumn{5}{ c| }{$T_1$ (ms)} & $\Gamma_s$ (s$^{-1}$) \\ \cline{2-6}
& 100 K & 150 K & 200 K & 250 K & 300 K &\\
\hline \hline
50 nm 	& $0.581 \pm 0.339$ & $0.519 \pm 0.341$ & $0.382 \pm 0.080$ & $0.132 \pm 0.034$ & $0.320 \pm 0.020$ & $2430 \pm 650$ \\
100 nm 	& $1.74 \pm 0.12$ & $1.68 \pm 0.18$ & $1.19 \pm 0.23$ & $1.02 \pm 0.15$ & $0.668 \pm 0.141$ & $587 \pm 63$ \\
250 nm 	& $25.6 \pm 1.2$ & $19.3 \pm 0.7$ & $7.80 \pm 0.20$ & $2.14 \pm 0.07$ & $1.26 \pm 0.03$ & $34.0 \pm 16.9$ \\
550 nm 	& $84.9 \pm 7.4$ & $42.5 \pm 2.2$ & $10.3 \pm 0.4$ & $2.58 \pm 0.03$ & $1.35 \pm 0.03$ & $8.12 \pm 22.03$ \\
10 $\mu$m  & $1200 \pm 580$ & $62.2 \pm 5.0$ & $12.0 \pm 0.4$ & $3.15 \pm 0.06$ & $1.36 \pm 0.03$ & ---\\
\hline
Annealed (5h) & $2.46 \pm 0.50$ & $1.41 \pm 0.26$ & $1.31 \pm 0.35$ & $0.885 \pm 0.188$ & $0.679 \pm 0.240$ & $531 \pm 217$ \\
Annealed (7h) & $4.37 \pm 1.37$ & $1.86 \pm 0.48$ & $1.37 \pm 0.57$ & $1.06 \pm 0.27$ & $0.847 \pm 0.595$ & $325 \pm 217$ \\
\hline
\end{tabular}
\end{center}
\end{table}
Next, we measured the spin relaxation times ($T_1$ and $T_2$) of the ND samples.
The experimental results of the 50-nm ND sample is shown in Fig.~\ref{fig:T1}(a).
The measurements of the $T_{1}$ and $T_2$ relaxation times of P1 centers were carried out using the inversion recovery and the spin echo  sequences, respectively.
The $T_1$ and $T_2$ measurements of P1 centers were performed at a microwave frequency of 115 GHz and 4.103 Tesla, corresponding to the center peak of the P1 EPR spectrum.
By fitting the change of the spin echo intensity with a single exponential function, we obtained $T_1$ to be $0.382\pm0.080$ ms, and $T_2$ is $0.413\pm0.007$ $\mu$s as shown in the inset of Fig.~\ref{fig:T1}(a) (see Supplementary Material for the description of the $T_1$ and $T_2$ determination).
Moreover, we measured temperature dependence of $T_1$ and $T_2$.
Figure~\ref{fig:T1}~(b) and Table~\ref{tab:T1} summarize the result of the $T_{1}$ measurements as a function of temperature.
We observed that $T_{1}$ times increase drastically by decreasing temperature.
In addition, the temperature dependence is strongly correlated with the size of the diamond powder.
To understand the temperature dependence, we first considered a contribution of the spin-lattice relaxation observed in bulk diamond.
According to the previous studies of $T_1$ of bulk diamond, the temperature dependence of $T_1$ is well explained by a spin-orbit induced tunneling model,~\cite{Reynhardt1998, Takahashi08} in which a spin flip event occurs due to the tunneling between P1's molecular axis orientations.
Using the spin-orbit induced tunneling model, we write that
$1/T_1$ is proportional to $T^5$, namely, $1/T_1 = C T^5$, where the $T$-linear term in the spin-orbit induced tunneling model~\cite{Reynhardt1998} is omitted because of its negligible contribution in the present temperature range.

By fitting the experimental data of the 10-$\mu$m diamond to the $T^5$ model, we obtained $C = (2.96 \pm 0.52) \times 10^{-10}$ $s^{-1}K^{-5}$, which is in a good agreement with previous finding.~\cite{Takahashi08, Reynhardt1998}
The result was obtained from a weighted fit analysis in order to take into account the uncertainty in $T_1$ values (See Supplementary Material for the details).
Furthermore, in cases of smaller diamond samples (from 550-nm to 50-nm NDs in Fig.~\ref{fig:T1}(b)), we observed a strong deviation from the $T^{5}$ model and found that $1/T_{1}$ at low temperatures highly correlates with the size of NDs.
Recent investigation of shallow NV centers as well as NV centers in a single ND showed that $T_{1}$ in NDs is attributed to surface spins.~\cite{Tetienne2013, Ofori2012, Kaufmann2013, Rosskopf2014}
Since the surface spins were also detected from the same ND samples in our experiment, it is likely that the surface spins also influence $T_1$ of P1 centers in NDs.
In order to take into account relaxation processes from both the surface spins and the spin-orbit induced tunneling, we consider the following for $1/T_{1}$,
\begin{equation}
\label{rate}
\frac{1}{T_{1}} = CT^{5}+\Gamma_{s},
\end{equation}
where $\Gamma_{s}$ is the $1/T_1$ contribution from surface spins, originated by fluctuations of the magnetic dipole fields from the surface spins.
$\Gamma_{s}$ is assumed to be independent of temperature in a temperature range of the present experiment.
In this $1/T_1$ model, when temperature increases, the first term (the spin-orbit induced tunneling contribution) increases.
Therefore, when a sample has a significant contribution from the surface spin relaxation, $1/T_1$ has less pronounced temperature dependence.
We performed a weighted fit analysis on 50-nm, 100-nm, 250-nm and 550-nm ND samples to determine their $\Gamma_{s}$ (See Supplementary Material for the details).
As shown in Fig~\ref{fig:T1} (b), we found a good agreement between the temperature dependence of $1/T_1$ and the model.
For example, we obtained that $\Gamma_{s}$ of the 50-nm ND sample was 2430 $\pm$ 650 $s^{-1}$.

\begin{table}[htb!]
\caption{\label{tab:T2} Summary of $T_2$ analyses. $T_2$ and $\overline{T_2}$ are represented by three significant figures.
The errors in $T_2$ represent the standard error of the mean. The errors in $\overline{T_2}$ were calculated as the 95\% confidence interval.}
\begin{center}
\begin{tabular} {|c |c |c |c |c |c |c | }
\hline
Sample & \multicolumn{5}{ c| }{ $T_2$ ($\mu$s)} & $\overline{T_2}$ ($\mu$s) \\ \cline{2-6}
& 100 K & 150 K & 200 K & 250 K & 300 K &\\
\hline \hline
50 nm & $0.518 \pm 0.002$ & $0.440 \pm 0.002$ & $0.413 \pm 0.007$ & $0.398 \pm 0.008$ & $0.211 \pm 0.036$ &  $0.474 \pm 0.060$ \\
100 nm & $0.756 \pm 0.007$ & $0.871 \pm 0.011$ & $0.835\pm 0.017$ & $0.942 \pm 0.020$ & $1.15 \pm 0.01$ &  $0.882 \pm 0.214$ \\
250 nm & $1.48 \pm 0.01$ & $1.42 \pm 0.01$ & $1.36 \pm 0.01$ & $1.41 \pm 0.01$ & $1.02 \pm 0.01$ &  $1.34 \pm 0.23$ \\
550 nm 	& $1.89 \pm 0.02$ & $2.03 \pm 0.01$ & $2.00 \pm 0.01$ & $2.15 \pm 0.01$ & $1.98 \pm 0.01$ &  $2.03 \pm 0.10$ \\
10 $\mu$m 	& $1.93 \pm 0.02$ & $1.84 \pm 0.02$ & $1.72 \pm 0.01$ & $1.59 \pm 0.01$ & $1.53 \pm 0.01$ & $1.65 \pm 0.17$ \\
\hline
Annealed (5hr)	& $0.510 \pm 0.091$ & $0.957 \pm 0.208$ & $0.341 \pm 0.252$ & $0.551 \pm 0.208$ & $0.869 \pm 0.095$ & $0.675 \pm 0.274$ \\
Annealed (7hr)	& $0.520 \pm 0.041$ & $0.509 \pm 0.036$ & $0.634 \pm 0.051$ & $0.604 \pm 0.015$ & $0.600 \pm 0.023$ & $0.589 \pm 0.048$ \\
\hline
\end{tabular}
\end{center}
\end{table}
Furthermore, we investigated the temperature- and size-dependence of $T_{2}$ relaxation time of P1 centers.
In contrast to the result of $T_{1}$, $T_{2}$ of P1 centers in the studied NDs does not show noticeable temperature dependence (see Fig.~\ref{fig:T1}(c)).
Table~\ref{tab:T2} shows the summary of the temperature- and size-dependence of $T_2$ as well as the mean $T_2$ ($\overline{T_2}$) which was obtained from a weighted fit analysis in order to take into account the errors in the $T_2$ values (See Supplementary Material for the details of the $\overline{T_2}$ analysis).
$\overline{T_2}$ for 50-nm ND and 550-nm samples were $0.474\pm0.060$ $\mu$s and $2.03\pm0.10$ $\mu$s, respectively.
Therefore, $T_2$ of the 50-nm NDs is approximately 4.3 times shorter than that of 550-nm NDs.
The result indicates the effect of the surface spins on $T_2$.
On the other hand, $T_2$ of 550-nm and 10-$\mu$m are similar ($\sim$2 $\mu$s).
This is probably because couplings to neighboring P1 centers dominates their $T_2$ processes.

\begin{figure}[hbt!]
\begin{center}
\includegraphics [width=45mm]{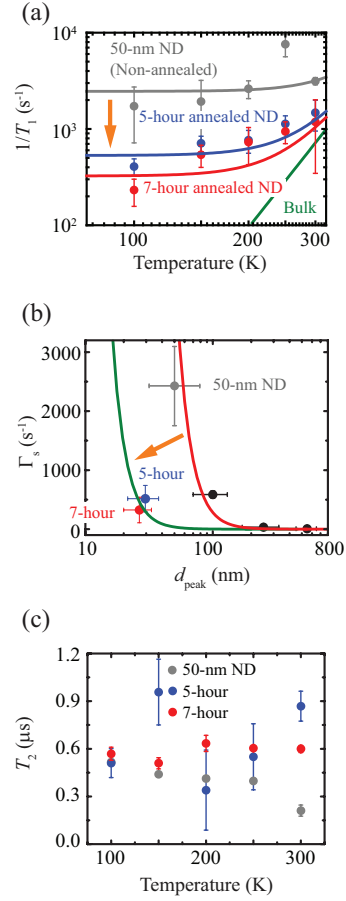}
\caption{\label{fig:T1en}
Temperature dependence of $T_1$ and $T_2$ of P1 centers in the annealed ND samples (initial diameter = 50 nm).
(a) $T_{1}$ of the annealed ND samples as a function of temperature.
Experimental data points are indicated by blue circles and red triangles for air annealing at 550 $^\circ$C for 5 hours and 7 hours, respectively.
The blue and red solid lines are corresponding fits utilizing Eq.~(\ref{rate}). $T_1$ data with no annealing (gray solid line) and the data of a bulk diamond (green solid line) are shown.
The arrow represents the reduction of $\Gamma_s$.
(b) $\Gamma_s$ as a function of the ND diameter.
The red solid line shows the fit result to the $\rho_s/d^4$ model for NDs without annealing.
The green solid line shows the $\rho_s/d^4$ line simulated for the annealed NDs.
The orange arrow represents the reduction of $\Gamma_s$.
The error bar in the $\Gamma_s$ is included and obtained by computing the 95 $\%$ confidence interval.
(c) $T_2$ as a function of temperature for the non-annealed and annealed samples.
}
\end{center}
\end{figure}
Finally, we study the spin relaxation times ($T_1$ and $T_2$) of the annealed NDs.
As shown in Fig.~\ref{fig:T1en}(a), $T_1$ times in the annealed diamond became longer after the annealing in the measured temperature range.
In addition, as shown in Fig.~\ref{fig:T1en}(a), the $T_1$ times of the annealed NDs are still shorter than that of bulk diamond, implying the existence of remaining surface spins.
To extract the contribution of the surface spins, we employed Eq.~(\ref{rate}) to determine $\Gamma_{s}$.
From the analysis, we indeed found that $\Gamma_{s}$ in the annealed NDs are smaller than that of the non-annealed samples.
The obtained $\Gamma_{s}$ are 531 $\pm$ 217 $s^{-1}$ and 325 $\pm$ 217 $s^{-1}$ for the NDs annealed at $550 ^{\circ}$C for 5 hours and 7 hours, respectively, which are 4.6 $\pm$ 2.2 and 7.5 $\pm$ 5.4 times smaller than that of the non-annealed 50-nm NDs as shown in Fig.~\ref{fig:T1en}(b) (see Supplementary Material for the calculation of the $\Gamma_{s}$ improvement factor).

We next discuss a model of the surface spin-induced $T_1$ ($\Gamma_{s}$).
As reported previously,~\cite{Steinert2013, Tetienne2013, Rosskopf2014} by considering fluctuating magnetic fields ($B_{dip}$) from surface spins, $\Gamma_{s}$ is proportional to the variance ($\langle B_{dip}^2 \rangle$) and the spin density ($\rho_{s}$).
By assuming that surface spins cover the whole surface uniformly, $B_{dip}^{2}(\overrightarrow{r}_{P1})\propto\int_{S}\rho_{s}b_{dip}^{2}(\overrightarrow{r}-\overrightarrow{r}_{P1}) dS$, where the radius vectors ($\overrightarrow{r}$ and $\overrightarrow{r}_{P1}$) define the locations of the surface and P1 spins relative to the center of the ND, respectively.
$b_{dip}(\overrightarrow{r}$) is the magnetic dipole field from the surface spins.
By taking into account the quantization axis of P1 and the surface spins along the external magnetic field and considering a spherical shape of NDs and a spatially uniform $\rho_s$, $b_{dip}^{2}(\overrightarrow{r})$ is proportional to $1/d^6$ and the surface integral is proportional to $d^2$, where $d$ is a diameter of a ND, the magnetic field fluctuations ($B^{2}_{dip}(\overrightarrow{r})$) is therefore proportional to $\rho_{s}/d^{4}$ and $\Gamma_{s}$ is also proportional to $\rho_{s}/d^{4}$.
The $\Gamma_{s}$ values obtained from the temperature dependence $T_1$ in Fig.~\ref{fig:T1}(b) were plotted as a function of the ND size in Fig.~\ref{fig:T1en}(b).
We found a good agreement between the obtained $\Gamma_{s}$ value and the  $1/d^{4}$ size dependence.
Thus, the result supports the $T_{1}$ relaxation mechanism in NDs due to the surface spins.
Furthermore, as shown in Fig.~\ref{fig:T1en}(b), $\Gamma_{s}$ of the annealed NDs are very different from the $1/d^4$ line of the non-annealed NDs, indicating significant reduction of the surface spin density.
Using the same model ($\Gamma_s \propto \rho_s/d^{4}$), we estimated that $\rho_s$ for the annealed NDs is $\sim$100 times smaller than that of the non-annealed NDs (Fig.~\ref{fig:T1en}(b)).

In addition the $T_2$ of the annealed diamond was studied.
As shown in Fig.~\ref{fig:T1en}(c), similarly to the non-annealed NDs, $T_2$ of the annealed diamond showed no temperature dependence.
The mean $T_2$ times were $0.675\pm 0.274$ $\mu$s and $0.589 \pm 0.048$ $\mu$s after the 5 and 7 hour annealing, respectively, showing that the extension of $T_2$ by a factor of $1.4 \pm 0.6$ and $1.2 \pm 0.2$, respectively (see Supplementary Material for the calculation of the $T_2$ improvement factor).
This improvement is due to the reduction of the surface spins.
The observed $T_2$ improvement is comparable with the previously reported result.~\cite{Tsukahara2019}
By considering the $T_2$ results of the non-annealed NDs, we speculate that the $T_2$ relaxation in the annealed NDs is caused by couplings to residual surface spins and P1 centers.

\section{Summary}
In summary, we investigated the relationship between the surface spins and the spin relaxation times ($T_1$ and $T_2$) of P1 centers in NDs.
We reduced the amount of the surface spins using air annealing.
The amount of the surface spins was characterized by HF EPR analysis.
The pulsed HF EPR experiment extracted the contribution of the surface spins on the $T_1$ relaxation successfully.
We found clear correlation between the amount of the surface spins and $T_1$.
In addition, the present study showed the improvement of $T_1$ and $T_2$ by removing the surface spins.
The finding of the present investigation sets the basis to suppress the spin relaxation process due to the surface spins in NDs which is critical for NV-based sensing applications.
The present method is also potentially applicable to improve spin and optical properties of other nanomaterials.

%
% Supplementary Materials
%
\section{Supplementary Material}
See the supplementary material for the $T_1$ and $T_2$ determination method, the EPR spectral analysis and the analyses of the temperature- and size-dependent $T_1$ and $T_2$.

%
% Acknowledgements
%
\begin{acknowledgments}
We thank Benjamin Fortman for useful discussion of the EPR data analyses.
This work was supported by the National Science Foundation (DMR-1508661 and CHE-1611134), the USC Anton B. Burg Foundation and the Searle scholars program. This material is also based upon work supported by the Chemical Measurement and Imaging program in the National Science Foundation Division of Chemistry under Grant No. CHE-2004252 (with partial co-funding from the Quantum Information Science program in the Division of Physics) (ST).

\end{acknowledgments}

\section*{DATA AVAILABILITY STATEMENT}
The data that support the findings of this study are available from the corresponding author
upon reasonable request.
%\bibliography{ZailiResearch}

\end{document}